\documentstyle{elsart}
\def\openone{\leavevmode\hbox{\small1\kern-3.3pt\normalsize1}}
\begin{document}
\begin{frontmatter}
\title{Universality of quantum Brownian motion} 
\author{Eric Lutz \thanksref{c}} and \author{Hans A. Weidenm\"uller} 
\address{Max Planck Institut f\"ur Kernphysik, Postfach 103980,
69029 Heidelberg, Germany}  
\thanks[c]{Corresponding Author. Fax: +49 6221 516 605; e-mail: eric.lutz@mpi-hd.mpg.de}
\begin{abstract}
Are Markovian master equations for quantum Brownian motion independent
of model assumptions used in the derivation and, thus, universal? With
the aim of answering this question, we use a random band--matrix
model for the system--bath interaction to derive Markovian master
equations for the time evolution of one--dimensional quantum systems
weakly coupled to a heat bath. We study in detail two simple systems,
the harmonic oscillator and the two--level system. Our results are in
complete agreement with those of earlier models, like the
Caldeira--Legget model and, in the large--band limit, with the Agarwal
equations (both with and without rotating--wave approximation). This
proves the universality of these master equations.
\end{abstract}
\begin{keyword}
{Quantum Brownian Motion; Random-matrix Theory; Universality}
\PACS{05.40.+j}
\end{keyword}
\end{frontmatter}

\section{Introduction}

The description of the interaction of an open quantum system with its
environment is an important problem in quantum physics. If the
environment is modeled as a heat bath, the interaction will lead to
relaxation and dissipation processes in the quantum system, and to an
irreversible approach toward equilibrium \cite{weiss}. During the last
decades, various models of this type have been introduced in different
branches of physics and chemistry. Notable examples are the Redfield
theory in nuclear magnetic resonance \cite{redfield}, the
Oppenheim--Romero-Rochin model in condensed--phase chemical physics
\cite{oppen}, the phase--space approach of Agarwal in quantum optics
\cite{agarwal1,agarwal2}, and the influence functional method used by
Caldeira and Leggett in condensed--matter physics \cite{caldeira}. The
Markovian master equations obtained in these approaches have been
recently compared in Ref.~\cite{kohen}.

The standard model for quantum dissipation (the Caldeira--Leggett
model) consists of a quantum system coupled to an infinite set of
harmonic oscillators. Caldeira and Leggett have shown that it is
always possible to treat the bath as an ensemble of independent
oscillators provided the system--bath coupling is weak. These authors
also assumed that the coupling is linear in both, the position
coordinate of the quantum system and the bath variables. For the
quantum system, this choice follows from the requirement that in the
classical limit, the friction force should be linear in the velocity.
For the bath, the choice was primarily made for computational
convenience. To the best of our knowledge, there is no compelling
argument for choosing the interaction term linear in the bath
coordinates.

It is expected, of course, that the relaxation process described by a
master equation should be insensitive to the detailed form of the
system--bath interaction. The work reported in the present paper aims
at proving this statement. We do so with the help of an alternative
model for the interaction. We use an ensemble of random matrices. The
ensemble encompasses all forms of system--bath interaction which are
linear in the position coordinate of the quantum system, and which
respect fundamental symmetries of the problem like time--reversal
invariance. The ensemble is characterized by a few parameters which
establish the relevant time scales. The Markovian master equations
derived in this fashion are then valid for all possible forms of the
interaction between quantum system and heat bath, except for a set of
measure zero. We find that in the the high--temperature limit, the
Markovian master equations derived by Caldeira and Leggett and others
are independent of both the specific structure of the bath and of the
specific form of the system--bath interaction. This establishes the
universality referred to in the title of our paper.

Random--matrix theory (RMT) was originally introduced by Wigner to
describe spectral fluctuations of quantum many--body systems such as
nuclei and has since been applied successfully in a wide range of
other fields such as quantum chaos and disordered mesoscopic systems
\cite{guhr}. To the best of our knowledge, a random--matrix approach
to relaxation has been first developed in nuclear physics in the
context of deeply inelastic heavy--ion collisions \cite{ko}. 
Since then, related models have been used to describe relaxation of a
non--degenerate two--level system \cite{pereyra}, dissipation in
complex quantum systems \cite{wilkinson,bulgac1} and, more recently,
the dynamics of a simple quantum system in a complex environment
\cite{bulgac2}.

A second motivation for our work relates to the use of random--matrix
models in {\em closed} quantum systems with many degrees of freedom. 
In such systems, only few degrees of freedom usually command physical
interest. We refer to such degrees of freedom (to the remainder) as to
the collective (the remaining) degree(s) of freedom, respectively. A
case in point is nuclear fission. Here, interest is focussed on the
shape degree of freedom, and little attention is usually paid to the
intrinsic degrees of freedom of the fissioning nucleus. In cases like
this, the dynamical behavior of the remaining degrees of freedom
depends, however, strongly on that of the collective degree of freedom
and, therefore, cannot be modeled as a heat bath. The success of RMT
in self--bound many--body quantum systems \cite{guhr} then suggests
that we model the remaining degrees of freedom in terms of a suitable
random--matrix model. We have in mind an ensemble of random matrices
which depends parametrically on the collective degree(s) of freedom.
Such an approach has been taken in the papers by Bulgac et
al. \cite{bulgac1}. Before investigating the consequences of such an
idea, it is necessary to study the limiting case where the environment
can indeed be modeled as a heat bath, and to ask whether in this case,
the heat bath can be replaced by a suitable random--matrix model. Our
proof of universality answers this question affirmatively.

The paper is organized as follows. In Section~\ref{mod}, we present
the model for system plus bath with emphasis on the statistical
properties of the interaction, a random band matrix. An approximate
form for the second moment (shown later to be equivalent to the
rotating--wave approximation) is introduced. In Section~\ref{der} we
derive the Markovian master equation for the averaged density operator
of the system. The range of validity of this equation is discussed in
Section~ \ref{dis}. In Sections~\ref{ap1} and \ref{ap2} we apply our
results to the damped harmonic oscillator and the dissipative
two--level system, respectively. We obtain a generalized band--width
dependent fluctuation--dissipation relation. We show that in the large
band--width limit we recover the Agarwal equations (with and without
rotating--wave approximation). Conclusions are drawn in Section~
\ref{con}.

\section{The model}
\label{mod}

We study the properties of a small quantum system $S$ coupled via a
random interaction to a large environment, considered as a heat bath. 
The Hamiltonian of the composite system is given by
\begin{equation}\label{eq1}
H = H_S\otimes \openone_S + \openone_B\otimes H_B  + Q \otimes V  =
H_0 + W 
\end{equation}
where $ H_S $  describes the system $S$ (for example, a particle
moving in a potential or a spin degree of freedom), $H_B$ describes
the bath $B$ (the actual form of $H_B$ is not specified) and $W =Q
\otimes V $  the system--bath interaction. We denote by $|n\rangle$
($|a\rangle$) the eigenstates of the system (bath) Hamiltonian with
eigenvalues $E_n$ ($\varepsilon_a$, respectively),
\begin{equation}
H_S|n\rangle = E_n|n\rangle \hspace{2cm} H_B |a\rangle= \varepsilon_a
|a\rangle \ .
\end{equation}
The product states $|na\rangle$ form a complete set for the composite
system. The operator $Q$ acts on the system $S$, and $V$ is a Gaussian
random band matrix acting on the bath. The first two moments of $V$
are given by
\begin{equation}
\label{eq2a}
\overline{V_{ab}} = 0 \hspace{1cm} \overline{V_{ab}V_{cd}}
 = (\delta_{ac}\delta_{bd} + \delta_{ad}\delta_{bc})
 \overline{{V_{ab}}^2}
\end{equation}
where the matrix $V_{ab}$ respects time--reversal symmetry and has
non--zero elements only in a band of width $\Delta$ along the
diagonal. More specifically, we adopt a form first given in
Ref.~\cite{ko}. This paper also contains a detailed justification
of the form of Eq.~(\ref{eq7}). This form has been widely used later,
cf. Ref~\cite{bulgac1,bulgac2}.
\begin{equation}
\label{eq7}
\overline{{V_{ab}}^2} = A_0 \left[ \rho(\varepsilon_a)
    \rho(\varepsilon_b)\right]^{-\frac{1}{2}}
    e^{-\frac{(\varepsilon_a - \varepsilon_b)^2}{2\Delta^2}} \ .
\end{equation}
Here $ A_0$ is the strength of the coupling, $\rho(\varepsilon)$ the
density of states of the bath, and $\Delta$ the band width. For $W$,
this implies
\begin{equation}
\label{eq2}
\mbox{(I)} \hspace{0.3cm} \overline{W_{ab}^{mn}} = 0 \hspace{0.5cm}
 \overline{W_{ab}^{mn}W_{cd}^{pq}}
 = (\delta_{ac}\delta_{bd} + \delta_{ad}\delta_{bc})
  Q_{mn} Q_{pq} \overline{{V_{ab}}^2} \ . 
\end{equation}
The form of the Hamiltonian (\ref{eq1}) is a generalization of the 
Hamiltonian considered in Ref.~\cite{pereyra}. There it was motivated
with the remark that in relaxation problems, the process is frequently
found to be insensitive to the details of the interaction. One may
therefore construct an ensemble of interactions and calculate the
average of the observable over this ensemble. We show below that this
is indeed the case.

In Eq.~(\ref{eq2}), only the part of the interaction acting on the
bath is modeled as a random matrix. This is physically sensible since
only the bath is supposed to be a complex system. We observe that as a
consequence, the variance (I) of $W$ has the inconvenient feature of 
being not symmetric in the variables of both the system and the bath. 
Therefore, we also consider a symmetrized form (II) of $W$ where the
entire interaction behaves as a random matrix. This form may be
thought of as an approximation to the full form (I). We show later
that form (II) leads to the rotating--wave approximation. It is given
by
\begin{equation}
\label{eq9}
\mbox{(II)} \hspace{0.3cm} \overline{W_{ab}^{mn}} = 0 \hspace{0.5cm}
\overline{W_{ab}^{mn}W_{cd}^{pq}} =
(\delta_{ac}\delta_{bd}\delta_{mp}\delta_{nq} +
\delta_{ad}\delta_{bc}\delta_{mq}\delta_{np}) |Q_{mn}|^2
\overline{{V_{ab}}^2} \ . 
\end{equation}

Let $\hat{\rho}$ be the density operator for system plus bath. The
von Neuman equation for $\hat{\rho}$ reads 
\begin{equation}\label{eq3}
\hat{\rho}(t) = U(t) \hat{\rho}(0)U^\dagger(t)
\end{equation}
where the time--evolution operator $U(t) = e^{-i Ht}$ obeys Dyson's
equation
\begin{equation}\label{eq4}
  U(t) = U_0(t) -  i \int_{0}^{t}  dt_1 U_0(t-t_1)W U(t_1) 
\end{equation}
and where $U_0(t) = e^{-i H_0t}$ denotes the free time--evolution
operator. (We put $\hbar = 1$ throughout). We define the reduced
operator for the system $S$ by $\hat{\rho}_S = tr_B[\hat{\rho}]$ where
the trace is taken over the bath states. We assume that the
interaction is turned on at $t = 0$ and that at that initial time $S$
and $B$ are not correlated. Then
\begin{equation}
  \hat{\rho}(0) = \hat{\rho}_S(0) \otimes \hat{\rho}_B(0)
\end{equation}
is the product of the initial density operators $\hat{\rho}_S(0)$ and
$\hat{\rho}_B(0)$ for system and bath, respectively. We suppose
further that at  all times $t \geq 0$ the bath is in thermal
equilibrium with temperature~$T$,
\begin{equation}\label{eq5}
  \hat{\rho}_B=  \frac{1}{Z} \sum_{a} e^{-\beta \varepsilon_a} |a
    \rangle \langle a| 
\end{equation}
where $Z$ is the canonical partition function and $\beta = (kT)^{-1}$.
Expression~(\ref{eq5}) can be shown (see Appendix~\ref{a1}) to be
equivalent to 
\begin{equation}\label{eq6}
  \hat{\rho}_B= |a^*\rangle \langle a^*| \hspace{1cm}
  \mbox{\underline{and}} \hspace{1cm} \rho(\varepsilon) = \rho_0
  e^{\beta \varepsilon} \ ,
\end{equation}
where the state $|a^* \rangle$ is defined by the temperature $T$.

\section{Derivation of the master equation}
\label{der}
In this Section we derive a Markovian master equation for
$\hat{\rho}_S(t)$. The equation applies provided the coupling between
system and bath is weak. More precisely, we use the following
assumptions.
\begin{itemize}
\item[i)] The time $t$ obeys the inequalities $t_\Delta \ll t
  \ll t_P$. Here $t_\Delta = 1/ \Delta$ is the duration time of a
  single action of the interaction and $t_P$ is the Poincare
  recurrence time of the system. This condition is always needed to
  describe a relaxation process in terms of a transport equation.
\item[ii)] For all states $|n \rangle$ and $|b \rangle$, the band
  width $\Delta$ has to satisfy the inequalities $\omega, \gamma \ll
  \Delta \ll kT$. Here $\gamma$ is the relaxation constant and
  $\omega$ denotes the mean level spacing of the system $S$. For the
  harmonic oscillator, $\gamma$ is defined in Eq.~(\ref{eq35}). For
  other systems, an analogous definition applies. Condition ii)
  requires weak coupling between bath and system and ensures the
  validity of the Markov approximation. It also requires the
  temperature $T$ to be larger than a minimum temperature $kT_m =
  \Delta$ and may, therefore, also be seen as defining a
  semiclassical approximation.
\end{itemize}
These assumptions are discussed in more detail in Section~\ref{dis}.

Because of the stochastic nature of the interaction $W$, the
time--evolution  operator~(\ref{eq4}) and, consequently, the density
operator~(\ref{eq3}) are themselves random variables. We have to
calculate their mean values. The averaging procedure consists in
expanding $U(t)$ in Eq.~(\ref{eq4}) in powers of $W$ (Born series),
averaging term by term, and finally summing up the whole series, see
Appendix~\ref{a2}. We always work in the limit in which the dimension
$N$ of the bath matrices tends to infinity. We consistently omit terms
of order $N^{-1}$ and smaller.

We illustrate the procedure by calculating the transition probability
per unit time. We suppose that $S$ is initially in some eigenstate $|m
\rangle$ with $\hat{\rho}_S(0) = |m \rangle \langle m| $, and we ask
for the probability to find the system in another state $|n \rangle$
at a later time $t$. We have
\begin{eqnarray}\label{eq10}
  P_n(t) = \langle n| \hat{\rho}_S(t) |n \rangle &=& \sum_b  \langle nb|
  \hat{\rho}(t) |nb \rangle \nonumber \\&=& \sum_{b} \langle nb  
  |U(t) |ma \rangle \langle ma | U^\dagger(t)|nb \rangle \ .  
\end{eqnarray}
Expanding $U(t)$ and $U^\dagger(t)$ up to first order in $W$ and
taking the average (denoted hereafter by a bar), we obtain
\begin{equation}
\overline{P_n(t)} = \sum_{b} \overline{\left| \langle nb| W |ma
      \rangle \right| ^2} \ 4 {\left( \frac{sin (E_n + \varepsilon_b -
      E_m - \varepsilon_a) \frac{t}{2}}{E_n + \varepsilon_b -
      E_m - \varepsilon_a}  \right)}^2 \ .
\end{equation}
For $t \gg t_\Delta$, the factor $4 (sin^2 \frac{1}{2}xt) / x^2$ is
sharply peaked at $x=0$ and may be approximated by $2\pi t
\delta(x)$. This yields
\begin{equation}
  \overline{P_n(t)} = 2 \pi t \sum_{b} \overline{\left| \langle nb| W
  |ma \rangle \right|^2} \ \delta(E_n + \varepsilon_b - E_m -
  \varepsilon_a) \ .
\end{equation}
The transition probability per unit time is defined as
\begin{equation}\label{eq11}
  W_{nm} = \frac{\overline{P_n(t)}}{t} = 2 \pi \sum_{b}
  \overline{\left| \langle nb| W |ma \rangle \right| ^2} \ \delta(E_n +
  \varepsilon_b - E_m - \varepsilon_a) \ .
\end{equation}
This is Fermi's Golden Rule.

We turn to the derivation of the master equation. The averaged density
operator $\overline{\rho}(t,t')$ for system plus bath obeys the
equation
\begin{eqnarray}
\label{eq15}
  \overline{\rho}(t,t') &=& \overline{U}(t) \hat{\rho}(0)
  \overline{U}^\dagger(t') \nonumber\\
  &+& \int_{0}^{t} d\tau \int_{0}^{t'} d\tau'
  \ \overline{U}(t-\tau) 
  \unitlength0.1cm
  \begin{picture}(23,0)\thinlines 
  \put(2,3.4){\line(0,1){2}}
  \put(2,5.4){\line(1,0){18}}
  \put(0,0){$W$}
  \put(5,0){$\overline{\rho}(\tau,\tau')$}
  \put(20,3.4){\line(0,1){2}}
  \put(18,0){$W$}
  \end{picture}   
   \overline{U}^\dagger(t'-\tau') \ .
\end{eqnarray}
The averaged time--evolution operator $\overline{U}(t)$ obeys
\begin{equation}
\label{eq16}
  \overline{U}(t) = {U}_0 (t) + \int_{0}^{t}dt_1 
  \int_{0}^{t_1} dt_2 \ {U}_0 (t-t_1)
  \unitlength0.1cm
  \begin{picture}(30,0)\thinlines 
  \put(2,3.4){\line(0,1){2}}
  \put(2,5.4){\line(1,0){25}}
  \put(0,0){$W$}
  \put(5,0){$U_0(t_1-t_2) $}
  \put(27,3.4){\line(0,1){2}}
  \put(25,0){$W$}
  \end{picture} 
   \overline{U}(t_2) \ .
\end{equation}
To obtain an evolution equation for the averaged reduced density
operator $\overline{\rho}_S$, we have to (i) solve
Eq.~(\ref{eq16}), (ii) substitute the resulting $\overline{U}(t)$ into
Eq.~(\ref{eq15}), (iii) take the trace over the bath degrees of
freedom and (iv) differentiate with respect to time. We shall see that
it is not always possible to perform step (i). This is the case, in
particular, for the unsymmetrized variance of Eq.~(\ref{eq2}).
Nevertheless, a master equation can still be derived in the weak
coupling limit. This is done in the next two Subsections.

\subsection{Case (II)}
\label{caseII}
We begin with the simpler case, i.e., with the approximate form
(II). In this case, the averaged time--evolution operator
$\overline{U}(t)$ is diagonal and reads (see Appendix~\ref{a2},
Eq.~(\ref{eqa8}))   
\begin{equation}
\label{eq18b}
  \overline{U}_{nb}(t) = e^{-i(E_n  + \varepsilon_b)t -
  \frac{\Gamma_{nb}}{2}t} \ .   
\end{equation}
The decay width $\Gamma_{nb}$ is given by Eq.~(\ref{eqa7}),
\begin{equation}
\label{eq18}
\label{width}
  \Gamma_{nb} = 2\pi \sum_{n_1b_1} \overline{\left| \langle nb| W |n_1b_1
  \rangle \right| ^2} \ \delta(E_n + \varepsilon_b -
  E_{n_1} - \varepsilon_{b_1}) \ .
\end{equation}
We substitute $\overline{U}(t)$ from Eq.~(\ref{eq16}) into
Eq.~(\ref{eq15}) and arrive at
\begin{eqnarray}
\label{eq18c}
&& \sum_b \langle nb |\overline{\rho}(t,t')| nb \rangle  =
  \sum_b e^{-i (E_n+\varepsilon_b) (t-t') - \frac{\Gamma_{nb}}{2}
  (t+t')} \langle nb|\hat{\rho}(0)|nb \rangle    \nonumber\\                 
& & \hspace{1cm}+ \int_{0}^{t} d\tau  \int_{0}^{t'} d\tau' 
  e^{(-i (E_n+\varepsilon_b) -\frac{\Gamma_{nb}}{2}) (t-\tau)} \nonumber\\
& & \hspace{1cm} \times \sum_{bn_1b_1}
  \overline{\left| \langle nb| W |n_1b_1
  \rangle \right| ^2}  \langle n_1b_1 |\overline{\rho}(\tau,\tau')| n_1b_1
  \rangle  e^{(i (E_n+\varepsilon_b) -\frac{\Gamma_{nb}}{2})
  (t'-\tau') } \ .
\end{eqnarray}
We take the time derivatives and get
\begin{eqnarray}
\label{master}
  \lefteqn{\left( \frac{\partial}{\partial t} + \frac{\partial}{\partial
    t'}\right) \sum_b \langle nb |\overline{\rho}(t,t')| nb \rangle =
-\sum_b \Gamma_{nb}  \langle nb |\overline{\rho}(t,t')| nb 
  \rangle} \nonumber \\
                  &+& \sum_{bn_1b_1}  \overline{\left| \langle nb| W |n_1b_1
  \rangle \right| ^2}  \int_{-t}^{0} dt_1 
  e^{(i (E_n+\varepsilon_b) +\frac{\Gamma_{nb}}{2}) t_1} \langle
  n_1b_1 |\overline{\rho}(t+t_1,t')| n_1b_1 
  \rangle \nonumber \\
                  &+& \sum_{bn_1b_1}  \overline{\left| \langle nb| W |n_1b_1
  \rangle \right| ^2}  \int_{-t'}^{0} dt_1' 
  e^{(-i (E_n+\varepsilon_b) +\frac{\Gamma_{nb}}{2}) t_1'} \langle n_1b_1
  |\overline{\rho}(t,t'+t_1')| n_1b_1 \rangle \ .
\end{eqnarray}
We have put $t_1 = \tau-t$ and  $t_1' = \tau'-t'$.

The right--hand side of Eq.~(\ref{master}) is easily interpreted. The
first term (the ``loss term'') corresponds to transitions which {\it
  deplete} the state $n$ whereas the two last terms (``gain terms'')
correspond to transitions which {\it feed} the state $n$. The gain
terms are non--local in time and, therefore, involve memory effects. 
In the limit of weak coupling, however, the process becomes Markovian
and the memory effects play no role \cite{weiden}. To see this, we
note that for short times (to zeroth order in W), Eq.~(\ref{eq15})
reduces to  
\begin{equation}\label{eq18a}
  \overline{\rho}(t,t') = U_0(t)  \hat{\rho}(0)
  {U_0}^\dagger(t') \ .
\end{equation}
We use this form in the gain terms and accordingly approximate
$\langle n_1 b_1|\overline{\rho}(t+t_1,t')| n_1b_1 \rangle $ by $e^{-i
  (E_{n_1}+\varepsilon_{b_1})t_1} \langle n_1b_1
|\overline{\rho}(t,t')| n_1b_1 \rangle$ and $\langle n_1b_1
|\overline{\rho}(t,t'+t_1')| n_1b_1 \rangle $ by $\langle n_1b_1
|\overline{\rho}(t,t')| n_1b_1 \rangle  e^{i
  (E_{n_1}+\varepsilon_{b_1})t_1'}$. Setting $ t=t'$, we obtain
\begin{eqnarray}
\label{eq19}
  &&\frac{d}{d t}\sum_b\langle nb |\overline{\rho}(t)| nb
  \rangle = -\sum_b \Gamma_{nb}  \langle nb |\overline{\rho}(t)| nb
  \rangle \nonumber \\
  &&+  \sum_{bn_1b_1}
   \overline{\left| \langle nb| W |n_1b_1
  \rangle \right| ^2} \langle n_1b_1 |\overline{\rho}(t)| n_1b_1
  \rangle \nonumber \\
  &&\times \left[ \int_{-t}^{0} dt_1  
  e^{(i( E_n+\varepsilon_b-E_{n_1}-\varepsilon_{b_1})
    +\frac{\Gamma_{nb}}{2}) t_1} + \int_{-t}^{0} dt_1 
  e^{(-i( E_n+\varepsilon_b-E_{n_1}-\varepsilon_{b_1})
    +\frac{\Gamma_{nb}}{2}) t_1} \right] \ .
\end{eqnarray}
Condition ii) implies $\Gamma_{nb} \ll E_n + \varepsilon_b -
E_{n_1} - \varepsilon_{b_1}$. Thus, we
neglect $\Gamma_{nb}$ in the integrands. This amounts to replacing
$\overline{U}(t)$ by $U_0(t)$. Moreover, for sufficiently large times
we have
\begin{equation}
\int_{-t}^t dt_1 e^{(i( E_n+\varepsilon_b-E_{n_1}-\varepsilon_{b_1})t_1}
\approx  2 \pi \delta(E_n \hspace{-0,1cm}+ \varepsilon_b \hspace{-0,1cm}-
      E_{n_1}\hspace{-0,1cm} - \varepsilon_{b_1}) \ .
\end{equation}
Hence, Eq.~(\ref{eq19}) reduces to 
\begin{eqnarray}
&&\frac{d}{dt} \sum_{b} \langle nb| \overline{\rho}(t) |nb \rangle 
= \nonumber \\
&& \hspace{1.3cm}2 \pi \sum_{bn_1b_1} \overline{\left| \langle nb| W |n_1b_1
\rangle \right| ^2} \delta(E_n \hspace{-0,1cm}+ \varepsilon_b
\hspace{-0,1cm} - E_{n_1}\hspace{-0,1cm} - \varepsilon_{b_1}) \langle
nb| \overline{\rho}(t) |nb \rangle \nonumber \\
&&\hspace{1cm}-  2 \pi  \hspace{-0,1cm}  \sum_{bn_1b_1} \overline{\left| \langle
nb| W |n_1b_1 \rangle \right| ^2} \delta(E_n \hspace{-0,1cm} +
\varepsilon_b \hspace{-0,1cm} - E_{n_1}\hspace{-0,1cm} -
\varepsilon_{b_1}) \langle n_1b_1| \overline{\rho}(t) |n_1b_1 \rangle
\ .
\end{eqnarray}
We have used Eq.~(\ref{eq18}) for $\Gamma_{nb}$. Using
Eq.~(\ref{eq10}) we finally obtain a master equation of the Pauli
type,
\begin{eqnarray}\label{eq20}
  \frac{d \overline{P}_n(t)}{dt} &=& \sum_{n_1}  W_{nn_1} \sum_{b_1}
  \langle n_1b_1| \overline{\rho}(t) |n_1b_1 \rangle - \sum_{n_1}
  W_{n_1n} \sum_{b} \langle nb| \overline{\rho}(t) |nb \rangle
  \nonumber \\
&=&  \sum_{n_1}  W_{nn_1}  \overline{P}_{n_1}(t) -  \sum_{n_1}
  W_{n_1n} \overline{P}_n(t) \ .
\end{eqnarray}
The transition probabilities are given by the Golden Rule expressions
(\ref{eq11}), 
\begin{eqnarray}\label{eq21}
W_{nn_1} &=& 2 \pi \sum_{b} \overline{\left| \langle nb| W |n_1b_1
      \rangle \right| ^2} \delta(E_n \hspace{-0,1cm} + \varepsilon_b
      \hspace{-0,1cm} - E_{n_1}\hspace{-0,1cm} - \varepsilon_{b_1}) \ ,
      \nonumber \\  
W_{n_1n} &=& 2 \pi \sum_{b_1} \overline{\left| \langle nb| W |n_1b_1
      \rangle \right| ^2} \delta(E_n \hspace{-0,1cm}+\varepsilon_b
      \hspace{-0,1cm} -E_{n_1}\hspace{-0,1cm} -\varepsilon_{b_1}) \ .
\end{eqnarray}
It may seem curious that in the evaluation of the gain term, it is
necessary to invoke the Markov approximation while in the loss term,
the limit of weak coupling apparently suffices. To explain this fact,
we recall that we always work in the limit of infinite matrix
dimension $N$. The loss term in the Pauli master equation is obtained
from single--side Wick contractions, symbolically written as
$\overline{VV}(\hspace{0,3cm}) \hspace{-0,1cm}:\hspace{-0,1cm}
(\hspace{0,3cm})$ or $(\hspace{0,3cm}) \hspace{-0,1cm}:\hspace{-0,1cm}
(\hspace{0,3cm} )\overline{VV}$, whereas the gain term is generated by
Wick contractions $\overline{V(\hspace{0,3cm})
  \hspace{-0,1cm}:\hspace{-0,1cm} (\hspace{0,3cm}) V}$ which connect
matrix elements in both amplitudes. Selection among the first type of
Wick contractions is affected by both, the limit $N \rightarrow
\infty$ and the weak--coupling limit. We exemplify this statement by
the terms of fourth order, given by the three Wick contractions
\unitlength0.1cm
\begin{picture}(12,0)\thinlines
\put(1.4,2.3){\line(0,1){2}}
\put(4.4,2.3){\line(0,1){2}}
\put(1.4,4.3){\line(1,0){3}}
\put(0,0){V}
\put(3,0){V}
\put(7.4,2.3){\line(0,1){2}}
\put(10.4,2.3){\line(0,1){2}}
\put(7.4,4.3){\line(1,0){3}}
\put(6,0){V}
\put(9,0){V}
\end{picture},
\unitlength0.1cm
\begin{picture}(12,0)\thinlines
\put(1.4,2.3){\line(0,1){2}}
\put(4.4,2.4){\line(0,1){1}}
\put(1.4,4.3){\line(1,0){9}}
\put(0,0){V}
\put(3,0){V}
\put(7.4,2.4){\line(0,1){1}}
\put(10.4,2.3){\line(0,1){2}}
\put(4.4,3.4){\line(1,0){3}}
\put(6,0){V}
\put(9,0){V}
\end{picture},
and
\unitlength0.1cm
\begin{picture}(12,0)\thinlines
\put(1.4,2.4){\line(0,1){1}}
\put(4.4,2.3){\line(0,1){2}}
\put(1.4,3.4){\line(1,0){6}}
\put(0,0){V}
\put(3,0){V}
\put(7.4,2.4){\line(0,1){1}}
\put(10.4,2.3){\line(0,1){2}}
\put(4.4,4.3){\line(1,0){6}}
\put(6,0){V}
\put(9,0){V}
\end{picture}.
The lines indicate which pairs of $V$'s are Wick--contracted. Of the
three, the last is neglected because $N \rightarrow \infty$ and the
second, because of weak coupling. In contradistinction, the form of
the gain terms in Eq.~(\ref{master}) is determined entirely by the
limit $N \rightarrow \infty$. Hence, an additional step is needed to
implement the weak--coupling limit.

\subsection{Case (I)}

For the non--symmetric form (\ref{eq2}) of the variance of the
interaction, the averaged time--evolution operator is not diagonal in
energy representation. Instead, we have
\begin{equation}
\langle nb|\overline{U}(t)|n'b' \rangle = \delta_{bb'} \langle
nb|\overline{U}(t)|n'b \rangle \ .
\end{equation}
Therefore, the matrix elements $\langle nb| \overline{U}(t) |n'b'
\rangle$ cannot be given in closed form (see Appendix~\ref{a2}).
However, the time derivative of these quantities can be obtained
explicitly in the limit of weak coupling. This suffices to obtain
the master equation. Aside from this difference, the derivation
proceeds in complete analogy to that given in the previous Subsection.
In analogy to Eq.~(\ref{eq18c}) we obtain
\begin{eqnarray}
&&\sum_b \langle nb |\overline{\rho}(t,t')| n'b \rangle  =
  \sum_{b{n_1}{n_2}} \langle nb| \overline{U}(t)|n_1b \rangle \langle
  n_1b| \hat{\rho}(0)| n_2 b \rangle \langle n_2b|
  \overline{U}^\dagger(t')| n'b \rangle \nonumber\\         
&& \hspace{3cm} \times \int_{0}^{t} d\tau  \int_{0}^{t'} d\tau' 
  \sum_{bn_1}\sum_{{n_2b_2}\atop{n_3n_4}}\langle nb|
  \overline{U}(t-\tau) |n_1b\rangle Q_{n_1n_2} Q_{n_3n_4}
  \overline{{V_{bb_2}}^2} \nonumber\\
&& \hspace{3cm}\times \langle n_2b_2 |\overline{\rho}
  (\tau,\tau')| n_3b_2 \rangle \langle n_4b|
  \overline{U}^\dagger(t'-\tau')|n'b \rangle \ .
\end{eqnarray}
We take the double time derivative, use Eqs.~(\ref{eqa29}) and
(\ref{eqa30}) for the time derivative of the time--evolution operator,
the Markov approximation in the gain terms, and the identity
\begin{equation}
\int_{0}^{\infty} d\tau e^{i x\tau} = i P\frac{1}{x} + \pi \delta(x) \ .
\end{equation}
We neglect the level shift due to the principal--value integral. All
this yields the master equation
\begin{eqnarray}
\label{eq27}
 &&\frac{d}{dt} \langle n|\overline{\rho}_S(t) |n' \rangle =  -i \langle n| 
  \left[H_S,\overline{\rho}_S(t) \right]|n' \rangle \nonumber \\
  &&\hspace{1cm}-  \frac{1}{2} \sum_{{n_1}{n_2}}
  W_{n{n_1}{n_1}{n_2}}^{(1)}\langle n_2|\overline{\rho}_S(t) |n'
  \rangle  - \frac{1}{2} \sum_{{n_1}{n_2}}
  W_{{n_1}n'{n_2}{n_1}}^{(2)}\langle n|\overline{\rho}_S(t) |n_2
  \rangle \nonumber \\
  && \hspace{1.3cm}\frac{1}{2} \sum_{{n_1}{n_2}}
  W_{n{n_1}{n_2}{n'}}^{(3)} \langle n_1|\overline{\rho}_S(t) |n_2
  \rangle + \frac{1}{2} \sum_{{n_1}{n_2}} W_{n{n_1}{n_2}{n'}}^{(4)}
  \langle n_1|\overline{\rho}_S(t) |n_2 \rangle \ .
\end{eqnarray}
We have defined the generalized transition probabilities
\begin{eqnarray}\label{eq28}
& &  W_{n{n_1}{n_1}{n_2}}^{(1)} = 2 \pi \sum_{{b_1}}
  Q_{n{n_1}}Q_{{n_1}{n_2}}  \overline{{V_{b{b_1}}}^2} \
  \delta(E_{n_2}\hspace{-0,1cm} + \varepsilon_b\hspace{-0,1cm} -
  E_{n_1} \hspace{-0,1cm}-\varepsilon_{b_1}) \ , \nonumber \\
& &  W_{{n_1}n'{n_2}{n_1}}^{(2)} = 2 \pi \sum_{{b_1}}
  Q_{{n_1}n'}Q_{{n_2}{n_1}}  \overline{{V_{b{b_1}}}^2} \
  \delta(E_{n_2}\hspace{-0,1cm} + \varepsilon_b\hspace{-0,1cm} -
  E_{n_1} \hspace{-0,1cm}-\varepsilon_{b_1}) \ , \nonumber \\
& & W_{n{n_1}{n_2}{n'}}^{(3)} = 2 \pi \sum_{b}
  Q_{n{n_1}}Q_{{n_2}{n'}}  \overline{{V_{b{b_1}}}^2} \
 \delta(E_{n_1}\hspace{-0,1cm} + \varepsilon_{b_1}\hspace{-0,1cm} -
  E_n \hspace{-0,1cm}-\varepsilon_b) \ , \nonumber \\
& & W_{n{n_1}{n_2}{n'}}^{(4)} = 2 \pi \sum_{b}
  Q_{n{n_1}}Q_{{n_2}{n'}}  \overline{{V_{b{b_1}}}^2} \
  \delta(E_{n_2}\hspace{-0,1cm} + \varepsilon_{b_1}\hspace{-0,1cm} -
  E_{n'} \hspace{-0,1cm}-\varepsilon_b) \ .
\end{eqnarray}

\section{Discussion: Time Scales}
\label{dis}

With an eye on the derivation given in the previous Section, we
discuss the various time scales appearing in our model. These time
scales play an essential role in defining the range of validity of the
master equation \cite{weiden,rau}.

According to the statistical {\it ansatz} in Eq.~(\ref{eq7}), the
interaction $V$ connects eigenstates of $H_B$ within an energy
interval $\sim \Delta$. Thus, the band width $\Delta$ can be
visualized as the average amount of energy exchanged during a single
action of $V$, and $t_{\Delta} = 1 /\Delta$ can be interpreted as
the duration time of a single action of $V$ (i.e., the time needed to
transfer the energy $\Delta$). A statistical
description in terms of a master equation can be valid only for times
$ t \gg t_\Delta$.

Any dynamical process in a finite--sized system will return (close) to
its initial state after a characteristic time, the Poincare recurrence
time $t_P$ (in the example of a two--level system, the Poincare time
corresponds to the Rabi period $t_P = 2\pi / (E_+ -E_-)$, where 
$ (E_+ -E_-) $ is the energy difference between the two (perturbed)
levels). When the bath is much larger than the system, the recurrence
time is essentially determined by the mean level spacing $D$ of the
bath, $t_P \sim 1 / D$. Obviously, $t_P$ tends to infinity with the
size of the bath. This is the condition of irreversibility. The
inequality $t \ll t_P$ must be fulfilled in order to have relaxation.

The weak--coupling condition ii) requires that the relaxation constant
$\gamma$ be much smaller than the amount $\Delta$ of energy
transferred during a single action of the interaction, $\gamma \ll
\Delta$ (see also Appendix~\ref{a2}). This condition has a simple
interpretation in terms of the times that correspond to these energies. 
The relaxation time $t_R = 1 /\gamma$ must be much larger than the
time $t_\Delta = 1/\Delta$ needed for a single action of $V$,
$t_\Delta \ll t_R$. 

There are two time scales which determine the memory time of the heat
bath \cite{aslangul,linden}: The time $t_\Delta$ (the inverse of the frequency cutoff of the
bath in the Caldeira-Leggett model) and the time $t_B = 1 /kT$. The
latter is purely quantum in origin. For high (low) temperature, thermal
(quantum) fluctuations dominate and the memory time of the bath is
given by $t_\Delta$ (by $t_B$, respectively). A crossover between
thermal and quantum fluctuations occurs at the crossover temperature
$kT_m = \Delta$. To garantee the validity of the Markov approximation,
the temperature must be much larger than the crossover temperature,
$\Delta \ll kT$. This condition can be rephrased in terms of length
scales: The range of the interaction must be much larger than the
thermal de Broglie wavelength of the Brownian particle $\lambda_{dB} =
1/\sqrt{4MkT}$.

\section{First Application: Harmonic oscillator}
\label{ap1}

We illustrate our results for the case where the system $S$ is a
harmonic oscillator with mass $M$ and frequency $\omega$. The
Hamiltonian $H_S$ is given by $H_S = \frac{p^2}{2M} + \frac{1}{2} M
\omega^2 x^2$, and the energy spectrum reads $E_n = (n+1/2) \ \omega,
\ n = 1,2,\ldots$. We assume a coupling linear in the position of the 
system, $Q = x$. We introduce the usual creation and annihilation
operators $a^{\dagger}$ and $a$. The elements of the matrix
$W_{ab}^{mn}$ vanish unless $|m - n| = 1$.

\subsection{Case (II)}

The transition probabilities in Eq.~(\ref{eq21}) are easily evaluated.
We use Eq.~(\ref{eq9}). For times $t\ll t_P$, we can replace the sum
over $b$ by an integral and obtain 
\begin{eqnarray}
   W_{n{n_1}} &=& 2 \pi  \sum_{b}  |Q_{nn_1}|^2
      \overline{{V_{bb_1}}^2} \delta(E_n \hspace{-0,1cm}+
      \varepsilon_b \hspace{-0,1cm} - E_{n_1} \hspace{-0,1cm} -
      \varepsilon_{n_1}) \nonumber \\ 
   &=& 2 \pi |Q_{nn_1}|^2 A_0 \int d\varepsilon_b \left[
     \frac{\rho(\varepsilon_b)}{\rho(\varepsilon_{b_1})}\right]
      ^\frac{1}{2} e^{-\frac{(\varepsilon_b - \varepsilon_{b_1})^2}
      {2\Delta^2}} \delta(E_n \hspace{-0,1cm} + \varepsilon_b
      \hspace{-0,1cm} - E_{n_1} \hspace{-0,1cm} - \varepsilon_{b_1})
      \nonumber \\
   &=& 2 \pi |Q_{nn_1}|^2 A_0 e^{\frac{\beta}{2} (E_{n_1}-E_n)}
      e^{-\frac{(E_{n_1}-E_n)^2}{2\Delta^2}} \ . 
\end{eqnarray}
In the last line, we have used Eq.~(\ref{eq6}) for the density of states
$\rho(\varepsilon)$. We use $Q = x = \sqrt{\frac{1} {2M\omega}} (a+
a^\dagger)$ and obtain for the only non--vanishing terms 
\begin{eqnarray}
  W_{nn-1} = W_0 n e^{-\frac{\beta}{2} \omega} \hspace{1cm} W_{nn+1} =
  W_0 (n+1) e^{\frac{\beta}{2} \omega} \nonumber \\
  W_{n-1n} = W_0 n e^{\frac{\beta}{2}  \omega} \hspace{1cm} W_{n+1n}=
  W_0 (n+1) e^{-\frac{\beta}{2} \omega}
\end{eqnarray}
where
\begin{equation}
\label{eq33}
  W_0 = \frac{A_0 \pi}{M\omega} e^{-\frac{\omega^2}{2\Delta^2}} \ .
\end{equation}
The master equation for the damped harmonic oscillator takes the form 
\begin{eqnarray}
  \dot{\overline{P}}_n(t) &=& 
  W_{nn-1} P_{n-1}(t) + W_{nn+1}P_{n+1}(t) 
  -(W_{n-1n}+W_{n+1n}) P_n(t) \nonumber \\
  &=&  W_0 e^{\frac{\beta}{2}\omega} \left[(n+1)P_{n+1}(t) - nP_n(t)
  \right] \nonumber \\
  &+& W_0 e^{-\frac{\beta}{2}\omega}\left[ nP_{n-1}(t)
  -(n+1)P_n(t) \right]
\end{eqnarray}
or, equivalently,
\begin{eqnarray}
\label{eq34}
  \dot{\overline{P}}_n(t) &=& 2\gamma (n_{th}+1)\left[(n+1)P_{n+1}(t) -
  nP_n(t) \right] \nonumber \\
&+& 2\gamma n_{th} \left[ nP_{n-1}(t) -(n+1)P_n(t)
  \right]
\end{eqnarray}
where we have defined the relaxation constant
\begin{equation}
\label{eq35}
  2\gamma = W_0 ( e^{\frac{\beta}{2}\omega} -
  e^{-\frac{\beta}{2}\omega}) \ ,
\end{equation}
and where $n_{th}$ denotes the average number of quanta at temperature
$T$,
\begin{equation}
\label{eq36}
n_{th} = \frac{1} {e^{\beta\omega}-1} \ . 
\end{equation}
Eq.~(\ref{eq34}) coincides with the Lax--Louisell master equation for
the harmonic oscillator in the energy representation and evaluated in
the Rotating Wave Approximation (RWA) \cite{louisell,agarwal1}
\begin{eqnarray}
\label{lax}
  \frac{d \overline{\rho}_S}{dt} =&-& i \omega [a^\dagger a,\overline{\rho}_S] + 
  \gamma(2 a \overline{\rho}_S a^\dagger - a^\dagger a \overline{\rho}_S -
  \overline{\rho}_S a^\dagger a)\nonumber \\
&+& 2\gamma n_{th} ( a \overline{\rho}_S a^\dagger + a^\dagger \overline{\rho}_S a 
- a^\dagger a \overline{\rho}_S - \overline{\rho}_S a a^\dagger) \ .
\end{eqnarray}
The RWA amounts to neglecting rapidly oscillating terms in the
interaction and is valid for weak damping, $\gamma \ll \omega$. The
coincidence between our result and the Lax--Louisell master equation
does not extend to the time dependence of the non--diagonal elements
$\langle n|\overline{\rho}_S | n'\rangle$ with $n \neq n'$ of the reduced
density operator. This is due to the form (\ref{eq9}) of the second
moment which supposes that the entire interaction acts as a random
matrix: In our approach, the gain term for these non--diagonal
elements vanishes because of the Kronecker delta for the states of the
system $S$ appearing in condition (II).

For the harmonic oscillator, Markovian master equations cannot simultaneously fulfill the following
three conditions: (i) The reduced density operator is positive
definite for all times $t > 0$; (ii) for $t \rightarrow \infty$, the
reduced density operator attains thermodynamic equilibrium; (iii) in
the classical limit, the equation is equivalent to a Langevin equation. 
This was shown in Ref.~\cite{lindblad1}. The master equation
(\ref{lax}) is of the Lindblad form \cite{lindblad} which guarantees
the positivity of the reduced density operator. As a consequence, the
quantum--classical correspondence with the Langevin equation is lost,
however.

\subsection{Case (I)}

We first evaluate the generalized transition probabilities of
Eq.~(\ref{eq28}). We proceed in complete analogy to case (II). We find,
for instance,
\begin{equation}
\label{eq37}
 W_{n{n_1}{n_1}{n_2}}^{(1)} = 2 \pi  A_0 Q_{n{n_1}}Q_{{n_1}{n_2}}
 e^{\frac{\beta}{2}(E_{n_2}-E_{n_1})}  e^{-\frac{(E_{n_2} -
 E_{n_1})^2} {2\Delta^2}} \ . 
\end{equation}
Using the explicit form of the matrix elements $Q$, we get
\begin{eqnarray}
&&W_{n{n-1}{n-1}n}^{(1)} = W_0 n e^{\frac{\beta\omega}{2}}
\hspace{1.8cm} W_{n{n+1}{n+1}{n+2}}^{(1)} = W_0 \sqrt{(n+1)(n+2)} \
e^{\frac{\beta\omega}{2}}  \nonumber \\   
&&W_{n{n+1}{n+1}{n}}^{(1)} = W_0 (n+1)
e^{-\frac{\beta\omega}{2}} \hspace{0.5cm}W_{n{n-1}{n-1}{n-2}}^{(1)} =
W_0 \sqrt{n(n-1)} e^{-\frac{\beta\omega}{2}}
 \end{eqnarray}
where $W_0$ is given by Eq.~(\ref{eq33}). Proceeding analogously for
the other transition probabilities and inserting the result into the
master equation, we obtain
\begin{eqnarray}
&& \frac{d}{dt} \langle n|\overline{\rho}_S(t) |n'
 \rangle = -i \omega (n-n') \langle n|\overline{\rho}_S(t) |n'
 \rangle) \nonumber \\
&&- \gamma n_{th} \left( \sqrt{n(n-1)}\, \langle
 n-2|\overline{\rho}_S |n' \rangle + (n+1) \langle n|\overline{\rho}_S
 |n' \rangle\right) \nonumber \\ 
&&- \gamma(n_{th}+1) \left(n \langle
 n|\overline{\rho}_S |n' \rangle + \sqrt{(n+1)(n+2)}\, \langle
 n+2|\overline{\rho}_S |n' \rangle\right ) \nonumber \\ 
&&- \gamma n_{th} \left( \sqrt{n'(n'+1)}\, \langle
 n|\overline{\rho}_S |n'-2 \rangle + (n'+1) \langle
 n|\overline{\rho}_S |n' \rangle\right) \nonumber \\
&&- \gamma(n_{th}+1) \left(n' \langle
 n|\overline{\rho}_S |n' \rangle + \sqrt{(n'+1)(n'+2)}\, \langle
 n|\overline{\rho}_S |n'+2 \rangle\right) \nonumber \\
&&+ \gamma  n_{th} \left(\sqrt{nn'} \langle
 n-1|\overline{\rho}_S |n'-1 \rangle + \sqrt{n(n'+1)} \langle n-1
 |\overline{\rho}_S |n'+1 \rangle\right) \nonumber \\
&&+ \gamma (n_{th}+1) \left(\sqrt{(n+1)n'} \langle n+1
 |\overline{\rho}_S |n'-1 \rangle \right) \nonumber \\
&&+ \gamma  n_{th} \left(\sqrt{nn'} \langle
 n-1|\overline{\rho}_S |n'-1 \rangle + \sqrt{(n+1)n'} \langle n+1
 |\overline{\rho}_S |n'-1 \rangle\right) \nonumber \\
&&+ \gamma (n_{th}+1) \left(\sqrt{n(n'+1)} \langle n-1
 |\overline{\rho}_S |n'\right) \nonumber \\
&& + 2 \gamma (n_{th}+1) \left(\sqrt{(n+1)(n'+1)} \langle n+1 |
  \overline{\rho}_S |n'+1 \rangle \right) 
\end{eqnarray}
It is easy to check that this result coincides with the master
equation in energy representation derived (without RWA) by Agarwal for
a harmonic oscillator linearly coupled to a bath,
\begin{eqnarray}
\label{eq38}
  \frac{d\overline{\rho}_S}{dt} & =& -i \omega[a^\dagger a,\overline{\rho}_S]
  \nonumber \\ 
&-&\gamma (a^\dagger a \overline{\rho}_S -2 a\overline{\rho}_Sa^\dagger +
\overline{\rho}_S a^\dagger a + a^2 \overline{\rho}_S - a\overline{\rho}_S a -
a^\dagger\overline{\rho}_Sa^\dagger +\overline{\rho}_S {a^\dagger}^2) \nonumber
  \\ 
&-& \gamma n_{th} \left( 2[a^\dagger,[a,\overline{\rho}_S]] +
[a^\dagger,[a^\dagger,\overline{\rho}_S]] + [a,[a,\overline{\rho}_S]] \right) \ .
\end{eqnarray}
The bath was modeled as an infinite set of harmonic oscillators, and
the projection operator technique was used \cite{agarwal1}.


\subsection{Fluctuation--Dissipation Theorem and High--Temperature
  Limit} 

With the help of Eq.~(\ref{eq33}), Eq.~(\ref{eq35}) for $\gamma$ takes
the form  
\begin{equation}
\label{eq45}
  \gamma = \frac{A_0 \pi}{M\omega} e^{-\frac{\omega^2}{2\Delta^2}}
  sinh\frac{\beta}{2}\omega \ .
\end{equation}
This equation expresses the fluctuation--dissipation relation. We see
that a small band width $\Delta$ tends to exponentially decrease the
damping coefficient. In view of our {\it ansatz} for the interaction
between system and bath, this fact is not surprising. In the limit of
large band width, $ \Delta \gg \omega$, Eq.~(\ref{eq45}) reduces to
the Agarwal fluctuation--dissipation relation \cite{agarwal1} between
the friction coefficient $\gamma$ and the diffusion coefficient $D$,
\begin{equation}
\label{eq46}
\gamma = \frac{D}{M\omega} e^{\frac{\beta\omega}{2}}
sinh\frac{\beta}{2}\omega \ .
\end{equation}
Here the diffusion constant $D$ is determined by the strength of the
coupling, 
\begin{equation}
  D = A_0 \pi  e^{-\frac{\beta\omega}{2}} \ .
\end{equation}
In the high--temperature limit $\beta \omega \ll 1$, Eq.~(\ref{eq46})
reduces to the Einstein relation 
\begin{equation}
  \gamma = \frac{D}{2MkT} \ .
\end{equation}
We also note that using $a = \left(\sqrt{\frac{M\omega}{2}} x + i
\sqrt{\frac{1}{2M\omega}}p \right)$ in Eq.~(\ref{eq38}) we find the
Caldeira--Leggett master equation \cite{caldeira}
\begin{equation}
   \frac{d\overline{\rho}_S}{dt}  = -i \omega \,[H_S,\overline{\rho}_S] -i
   \gamma \, [x,\{p,\overline{\rho}_S\}] - D \,[x,[x,\overline{\rho}_S] \ .
\end{equation}
This equation does not have the Lindblad form. This implies that 
the positivity of the reduced density operator can be violated for
certain initial states. In Ref.~\cite{nemes}, it was shown that
positivity is guaranteed provided the dispersion $\sigma_{xx} = <x^2>
- <x>^2$ of the initial wave packet obeys the condition $\sigma_{xx}
\geq {\lambda_{dB}}^2$.

\section{Second Application: Two--level system}
\label{ap2}

We derive the master equation for a two--level
system with upper (lower) level $|+\rangle$ ($|-\rangle$,
respectively). We introduce the Pauli spin matrices $\sigma_x$,
$\sigma_y$ and $\sigma_z$. The Hamiltonian of the system takes
the form
\begin{equation}
  H_S = \frac{1}{2} \omega_0 \sigma_z
\end{equation}
where $\omega_0$ is the energy separation between the two levels. We
write $E_n = \frac{1}{2} \omega_0  n, \ n=\pm1$ and take $Q =
\sigma_x$. For the generalized transition probabilies we find
\begin{eqnarray}
  W_{+--+}^{(1)} &=& W_0 e^{\frac{\beta}{2} \omega_0} \nonumber \\
  W_{-++-}^{(1)} &=& W_0 e^{-\frac{\beta}{2} \omega_0}
\end{eqnarray} 
and similar expressions for $ W^{(2)}$ , $ W^{(3)}$ and $W^{(4)}$. 
Here $W_0$ is given by Eq.~(\ref{eq33}) with $\omega$ replaced by
$\omega_0$ and $M$ set to unity.  Thus,
\begin{eqnarray}
  \frac{d}{dt} \langle+ | \overline{\rho}_S | + \rangle& =&
  -\frac{1}{2} W_0 e^{\frac{\beta}{2}  \omega_0} \langle+ |
  \overline{\rho}_S | + \rangle -\frac{1}{2}  W_0e^{\frac{\beta}{2}
    \omega_0} \langle+ | \overline{\rho}_S | + \rangle \nonumber \\
  && +\frac{1}{2}  e^{-\frac{\beta}{2}  \omega_0} \langle- |
  \overline{\rho}_S | - \rangle +\frac{1}{2}  e^{-\frac{\beta}{2}
    \omega_0} \langle- | \overline{\rho}_S | - \rangle \nonumber \\
  &= & -2 \gamma (n_{th}+1) \langle+ | \overline{\rho}_S | + \rangle
  +2 \gamma n_{th} \langle- | \overline{\rho}_S | - \rangle
\end{eqnarray}
and, in a similar way,
\begin{eqnarray}
  \frac{d}{dt} \langle- | \overline{\rho}_S | - \rangle& =& -2 \gamma
  n_{th} \langle- | \overline{\rho}_S | - \rangle 
  +2 \gamma (n_{th}+1) \langle+ | \overline{\rho}_S | + \rangle \ ,
  \nonumber \\
  \frac{d}{dt} \langle- | \overline{\rho}_S | + \rangle & =& i
  \omega_0  \langle- | \overline{\rho}_S | + \rangle - \gamma
  (2n_{th}+1) \left[\langle- | \overline{\rho}_S | + \rangle -\langle+
    | \overline{\rho}_S | -\rangle \right] \ , \nonumber \\ 
  \frac{d}{dt} \langle+ | \overline{\rho}_S | - \rangle & =& -i
  \omega_0  \langle+ | \overline{\rho}_S | - \rangle - \gamma
  (2n_{th}+1) \left[\langle+ | \overline{\rho}_S | - \rangle -\langle-
    | \overline{\rho}_S | +  \rangle \right] \ .
\end{eqnarray}
Here $n_{th}$ is given by Eq.~(\ref{eq36}) with $\omega$ replaced by
$\omega_0$. Simple manipulations with the spin matrices show that the
master equation for the two--level system may be written as
\begin{eqnarray}
  \frac{d}{dt} \overline{\rho}_S &=&- 2 \gamma (n_{th}+1)\,
  \sigma_+\sigma_- \overline{\rho}_S\sigma_+\sigma_- +2 \gamma
  n_{th} \,\sigma_+\overline{\rho}_S \sigma_-  \nonumber \\
 &&-  i \omega_0  \,\sigma_+\sigma_-\overline{\rho}_S  \sigma_-\sigma_+ -\gamma
  (2n_{th}+1) \left[ \sigma_+\sigma_-\overline{\rho}_S
    \sigma_-\sigma_+ - \sigma_+  \overline{\rho}_S   \sigma_+ \right]
  \nonumber \\
 &&+  i \omega_0  \,\sigma_-\sigma_+\overline{\rho}_S  \sigma_+\sigma_- -\gamma
  (2n_{th}+1) \left[ \sigma_-\sigma_+\overline{\rho}_S
    \sigma_+\sigma_- - \sigma_-  \overline{\rho}_S   \sigma_- \right]
  \nonumber \\
  &&- 2 \gamma n_{th}\,
  \sigma_-\sigma_+ \overline{\rho}_S\sigma_-\sigma_+ +2 \gamma
  (n_{th}+1)\, \sigma_-\overline{\rho}_S \sigma_+  
\end{eqnarray}
or, more simply,
\begin{eqnarray}\label{eq48}
 \frac{d}{dt} \overline{\rho}_S &=& -\frac{1}{2} \omega_0[\sigma_z,
 \overline{\rho}_S] \\
&+& \gamma n_{th} (2 \sigma_+ \overline{\rho}_S
 \sigma_- + \sigma_+ \overline{\rho}_S \sigma_+ + \sigma_-
 \overline{\rho}_S \sigma_- - \overline{\rho}_S\sigma_- \sigma_+ -
 \sigma_- \sigma_+\overline{\rho}_S) \nonumber \\
 &+& \gamma (n_{th}+1) ( \sigma_+ \overline{\rho}_S \sigma_+ +
 \sigma_- \overline{\rho}_S \sigma_- + 2 \sigma_-\overline{\rho}_S
 \sigma_+ - \sigma_+ \sigma_- \overline{\rho}_S - \overline{\rho}_S
 \sigma_+ \sigma_-) \ .\nonumber
\end{eqnarray}
We have defined the operators $ \sigma_- = |-\rangle \langle+| $ and $
\sigma_+ = |+\rangle \langle-| $. For $n_{th}=0$, Eq.~(\ref{eq48})
reduces to the Agarwal equation for spontaneous emission of a
two--level atom \cite{agarwal2}. It may seem surprising that
Eq.~(\ref{eq48}) applies for $T=0$ where the inequality $\Delta \ll T$
is clearly violated and non-Markovian effects are present. In
Ref.~\cite{unruh} it was shown that the high--temperature master
equation is obeyed even at low temperatures provided the time $t$ is
larger than the memory time of the bath, $t \gg t_B$. This implies
that condition ii) of Section (\ref{der}) may be replaced by the less
restrictive condition $t_B \ll t \ll t_P$.

\section{Conclusion}
\label{con}

Starting from a random band--matrix model for the system--bath
interaction, we have derived a Markovian master equation for the
average reduced density operator of a one--dimensional quantum system. 
We have assumed that the system--bath interaction is linear in the
position coordinate of the quantum system, and we have considered two
cases: (I) That part of the interaction which depends on the bath
variables is a member of an ensemble of random matrices of proper
symmetry; (II) the entire interaction is a member of an ensemble of
random matrices of proper symmetry. The form of the master equation
differs in both cases. The equation is valid in a domain of parameter
values specified by the inequalities i) and ii) of Section~\ref{der}.

We have applied the master equation to two cases, the damped harmonic
oscillator and the dissipative two--level system. For the damped
harmonic oscillator and case (I), we have obtained the same equation
as Argawal who considered a harmonic oscillator coupled to a heat bath
which was modeled as an infinite set of harmonic oscillators. In the
limit of high temperature, this equation coincides with the master
equation of the Caldeira--Leggett model. For case (II), our master
equation for the diagonal elements of the reduced density operator is
identical to the Lax--Louisell master equation evaluated in the
Rotating Wave Approximation (RWA). This is because in case (II) we
impose conditions upon the interaction matrix elements of the position
coordinate of the quantum system. These conditions are tantamount to
the RWA. For the non--diagonal elements of the reduced density
operator, these same conditions imply the vanishing of the gain
terms. In this point our result differs from that obtained by Lax
and Louisell. For the two--level system and case (I), our master
equation reduces, at $T = 0$, to the Argawal equation for spontaneous
emission of a two--level atom.

We conclude that Markovian master equations for quantum Brownian
motion derived in the weak coupling limit possess universal validity:
These equations are independent of the specific microscopic model used
for their derivation. This is not true of approximations like the
Rotating Wave Approximation. Typically, such approximations violate
certain invariance requirements (translational invariance in the case
of the RWA) and, thereby, lose universal validity.

\begin{appendix}

\section{Appendix}
\label{a1}

The equivalence of Eqs.~(\ref{eq5}) and (\ref{eq6}) is closely related
to the equivalence of the microcanonical and the canonical ensemble in
the thermodynamic limit $N \rightarrow \infty$ \cite{huang}. We thus
consider a thermodynamical system in contact with a heat bath. The
(canonical) partition function is given by
\begin{equation}
  Z(\beta) = \int_{0}^{\infty} d\varepsilon \rho(\varepsilon)
  e^{-\beta \varepsilon} = \int_{0}^{\infty} d\varepsilon  e^{-\beta
  \varepsilon + \ln\rho(\varepsilon)} \ .
\end{equation}
Since $\varepsilon$ and $ \ln\rho(\varepsilon)$ grow with $N$, the
integral can be evaluated by a saddle--point approximation. Expanding
the integrand up to second order around its maximum $\varepsilon^*$,
we get 
\begin{equation}\label{eqa1}
  Z(\beta) = \rho(\varepsilon^*) e^{-\beta \varepsilon^*}
  \int_{0}^{\infty} d\varepsilon\, \exp\left(\frac{1}{2} (\varepsilon -
  \varepsilon^*)^2 \left(\frac{\partial^2 \ln\rho(\varepsilon)}
  {\partial \varepsilon^2}\right)_{\varepsilon=\varepsilon^*} \right)
  \ .
\end{equation}
The distribution in energy is a Gaussian centered at $\varepsilon^*$
with a width
\begin{equation}
  \Delta \varepsilon = \sqrt{- \left(\frac{\partial^2
  \ln\rho(\varepsilon)}{\partial \varepsilon^2}\right)} = \sqrt{k T^2
  C_V} \ .
\end{equation}
We have used the fact that $k \,\ln\rho(\varepsilon)$ is the
microcanonical entropy. For $N \rightarrow \infty$, $\Delta
\varepsilon / \varepsilon^* \sim 1/\sqrt{N}$ becomes negligibly
small and the Gaussian approaches a $\delta$-function. Hence,
\begin{equation}
  \hat{\rho}_B(0)=  \frac{1}{Z} \sum_{a} e^{-\beta \varepsilon_a} |a
    \rangle \langle a| \simeq |a^*\rangle \langle a^*|
\end{equation} 
where $ |a^*\rangle $ is the eigenvector corresponding to the
eigenvalue $\varepsilon^*$. Moreover, according to Eq.~(\ref{eqa1})
the density of states can be approximated locally by
\begin{equation}
  \rho(\varepsilon^*) = \rho_0  e^{\beta \varepsilon^*} \ . 
\end{equation}

\section{Appendix}\label{a2}

\subsection{The average propagator}

The propagator $ K(t) $ is defined by 
\begin{equation}\label{eqa1a}
  K(t) = U(t) \theta(t) = e^{-iHt}\theta(t) 
\end{equation}
and obeys Dyson's equation
\begin{equation}\label{eqa2}
  K(t) = K_0(t) -i \int_{-\infty}^{\infty} dt_1 K_0(t-t_1)WK(t_1)  
\end{equation}
where $K_0(t) = e^{-iH_0t}\theta(t)$ is the free propagator and
$\theta(t)$ the unit step function.

To calculate the average of $ K(t) $, we use the energy representation
and introduce the following pair of Fourier transforms 
\begin{equation}
  G(E) = \frac{1}{i} \int_{-\infty}^{\infty} dt e^{iEt}K(t)\ ,
  \hspace{1cm} K(t) = \frac{1}{2\pi i}\int_{-\infty}^{\infty} dt
  e^{-iEt}G(E) \ . 
\end{equation}
The transforms of (\ref{eqa1a}) and  (\ref{eqa2}) are then given by
\begin{equation}
  G(E) = \frac{1}{E-H +i\varepsilon} \hspace{1cm} \varepsilon
  \rightarrow 0^+   
\end{equation}
and 
\begin{equation}
\label{eqa4}
  G = G_0 +G_0WG = \sum_{s=0}^{\infty} G_0(WG_0)^s \ , 
\end{equation}
respectively. 

To average $G$, we use Eqs.~(\ref{eq2},\ref{eq9}) and the rule that
the average over a product of Gaussian distributed $W$'s equals the
sum over all ways of Wick--contracting pairs of $W$'s. We show below that in the 
limit of weak coupling, only contractions between adjacent pairs of
$W$'s have to be taken into account. We find
\begin{eqnarray}\label{eqa5}
  \overline{G} = \sum_{s=0}^{\infty}\overline{ G_0(WG_0)^s} &=& G_0 + 
  G_0\overline{WG_0W}G_0 + G_0\overline{WG_0W}G_0\overline{WG_0W}G_0 +
  \cdots \nonumber \\  &=& G_0 \frac{1}{1-\overline{WG_0W}G_0} \ .
\end{eqnarray}

\subsubsection{Case (II)}

Since 
\begin{equation}
  \langle nb|G_0(E)|n'b'\rangle = \delta_{nn'}\delta_{bb'}
  \frac{1}{E-(E_n+\varepsilon_b) +i\varepsilon}=
  \delta_{nn'}\delta_{bb'}(G_0)_{nb}
\end{equation}
and 
\begin{eqnarray}
  \langle nb|\overline{WG_0W}|n'b'\rangle &=&
  \sum_{n_1b_1}\overline{W_{bb_1}^{nn_1}(G_0)_{n_1b_1}W_{b_1b'}^{n_1n'}} 
  (G_0)_{n'b'} \nonumber \\ &=& \delta_{nn'}\delta_{bb'} (G_0)_{nb}
  \sum_{n_1b_1} \overline{{W_{bb_1}^{nn_1}}^2}(G_0)_{n_1b_1} \ ,  
\end{eqnarray}
the matrix elements of $\overline{G}$ become
\begin{eqnarray}
  \langle nb|\overline{G}|n'b' \rangle &=&\frac{ \delta_{nn'}
  \delta_{bb'}}{E-(E_n+\varepsilon_b) -
  \sum_{n_1b_1}\overline{{W_{bb_1}^{nn_1}}^2} (G_0)_{n_1b_1}}
\nonumber \\&=& \frac{
  \delta_{nn'} \delta_{bb'}}{E-(E_n+\varepsilon_b)-R_{nb}(E)} 
\end{eqnarray}
which shows that $\overline{G}$ is diagonal in the unperturbed energy
basis. Here,
\begin{equation}
  R_{nb}(E) = \Delta_{nb}(E) -i \frac{\Gamma_{nb}(E)}{2}  
\end{equation}
where
\begin{eqnarray}
  \Delta_{nb}(E) &=& P\sum_{n_1b_1}\frac{\overline{{W_{bb_1}^{nn_1}}^2}}
  {E-(E_{n_1} + \varepsilon_{b_1})} \ , \nonumber \\
  \Gamma_{nb}(E) &=& 2 \pi \sum_{n_1b_1} \overline{{W_{bb_1}^{nn_1}}^2}
  \delta(E-(E_{n_1}+\varepsilon_{b_1})) \ .
\end{eqnarray}
For sufficiently high temperature, $\Gamma_{nb}(E)$ depends slowly on energy
 and can be approximated by 
\begin{eqnarray}\label{eqa7}
  \Gamma_{nb}(E)\simeq \Gamma_{nb}(E_n+\varepsilon_b)
  &=&\Gamma_{nb}\nonumber \\
  &=& 2 \pi \sum_{n_1b_1} \overline{{W_{bb_1}^{nn_1}}^2}
  \delta(E_n+\varepsilon_b-E_{n_1}-\varepsilon_{b_1}) \ .
\end{eqnarray}
For the averaged propagator, this yields
\begin{equation}
  \overline{G}_{nb}(E) = \frac{1}{E-(E_n +\varepsilon_b)+i
  \frac{\Gamma_{nb}}{2}}
\end{equation}
where we have neglected the level shift $\Delta_{nb}$. Transforming
back to time representation, we obtain
\begin{equation}\label{eqa8}
  \overline{K}_{nb}(t) = e^{-i(E_nt +\varepsilon_b)-
  \frac{\Gamma_{nb}}{2}t} \theta(t) \ .  
\end{equation}
In the limit of weak coupling, contractions between non--adjacent
pairs of W's are negligible. This can be shown by inspecting the
contributions from the various products of Wick contractions. We
consider here the simplest case and compare \cite{agassi} the
imaginary parts of
$\langle nb|$
\unitlength0.1cm
\begin{picture}(12,0)\thinlines
\put(2,2.8){\line(0,1){2}}
\put(11,2.8){\line(0,1){2}}
\put(2,4.8){\line(1,0){9}}
\put(0,0){$W$}
\put(4,0){$G_0$}
\put(9,0){$W$}
\end{picture}
$|n'b' \rangle$ and $\langle nb|$
\unitlength0.1cm
\begin{picture}(31,0)\thinlines
\put(2,2.8){\line(0,1){2}}
\put(11,3){\line(0,1){1}}
\put(2,4.8){\line(1,0){27}}
\put(0,0){$W$}
\put(4,0){$G_0$}
\put(9,0){$W$}
\put(20,3){\line(0,1){1}}
\put(29,2.8){\line(0,1){2}}
\put(11,4){\line(1,0){9}}
\put(13,0){$G_0$}
\put(18,0){$W$}
\put(22,0){$G_0$}
\put(27,0){$W$}
\end{picture} 
$|n'b' \rangle$ (the real parts  represent level shifts and do not
contribute to the decay width). We are thus led to compare
$\Gamma_{nb}(E)$ with \linebreak $ \sum_{n_1b_1} \overline{{W_{bb_1}^{nn_1}}^2}
(G_0)_{n_1b_1}^2\Gamma_{n_1b_1}(E)$. These quantities are calculated
as in Section~\ref{ap1}. Assuming that $|Q_{nn_1}|^2$ vanishes unless
the states $n_1$ and $n$ are close in energy, we find that in the
high--temperature limit $\omega \ll \Delta \ll kT$, $\Gamma_{nb}(E)$
may be replaced by  $\Gamma_{nb}(E_n+\varepsilon_b)$,
and that in the weak--coupling limit $\gamma \ll \Delta$, the second
term may be omitted in comparison with the first.

\subsubsection{Case(I)}

In this case, the propagator is not diagonal in the energy
representation and we have instead 
\begin{equation}
   \langle nb|\overline{G}|n'b' \rangle =\delta_{bb'} \langle nb |
   \overline{G}|n'b \rangle \ .
\end{equation}
As a consequence, the series (\ref{eqa5}) is not geometric any more
and cannot be summed easily. We thus have to find another method to
evaluate the average propagator. We actually calculate only the time
derivative of the average propagator which is sufficient for the
derivation of the master equation.

Iterating the Dyson equation (\ref{eqa4}) once
\begin{equation}
  G=G_0+G_0W+G_0WG_0WG 
\end{equation}
and taking the average
\begin{equation}
  \overline{G}=G_0+G_0\overline{W}+G_0\overline{WG_0WG} =  G_0 +
  G_0\overline{WG_0W} \hspace{0,1cm} \overline{G}
\end{equation}
yields the same result as in Eq.~(\ref{eqa5}). We use this fact to
average the time--evolution operator. In the interaction picture, we
have
\begin{equation}\label{eqa20}
  i \frac{d\tilde{U}(t)}{dt} = \tilde{W}(t) \tilde{U}(t)
  \hspace{2cm} \tilde{U}(0) = U(0) = 1  
\end{equation}
where 
\begin{equation}
  \tilde{U}(t) = e^{iH_0t} U(t) \hspace{1cm} \mbox{and} \hspace{1cm} 
  \tilde{W}(t) = e^{iH_0t} W e^{-iH_0t} \ .
\end{equation}
This equation can also be written in integral form
\begin{equation}\label{eqa21}
  \tilde{U}(t) = 1 - i \int_{0}^{t}  dt_1 \tilde{W}(t_1)
  \tilde{U}(t_1) \ . 
\end{equation}
Inserting Eq.~(\ref{eqa21}) into Eq.~(\ref{eqa20}) and averaging, we
find
\begin{eqnarray}
  &&\frac{d \overline{\langle nb|\tilde{U}(t)| nb\rangle}}{dt}
  =\nonumber \\
 &&- \int_{0}^{t}  dt_1 \sum_{{n_1b_1}\atop{n_2b_2}} \overline{ \langle
  nb|\tilde{W}(t)|n_1b_1 \rangle\langle
  n_1b_1|\tilde{W}(t_1)|n_2b_2\rangle} \hspace{0,1cm} \overline{ 
  \langle n_2b_2|\tilde{U}(t_1)| nb\rangle} \ .
\end{eqnarray}
We have $\overline{W_{bb_1}^{nn_1}W_{b_1b_2}^{n_1n_2}} = \delta_{bb_2}
Q_{nn_1} Q_{n_1n_2} \overline{{V_{bb_1}}^2}$, hence
\begin{eqnarray}
  \frac{d \overline{\langle nb|\tilde{U}(t)| n'b\rangle}}{dt} &=&
  - \int_{0}^{t}  d\tau \sum_{n_1b_1n_2}
  e^{i(E_n-E_{n_2})t}e^{i(E_{n_2}+\varepsilon_b -E_{n_1}-\varepsilon_{b_1})\tau}  \nonumber \\ 
& & \times Q_{nn_1}Q_{n_1n_2}\overline{{V_{bb_1}}^2}
  \hspace{0,1cm}\overline{ \langle n_2b|\tilde{U}(t- \tau)| n'b\rangle}   
\end{eqnarray}
where we have set $\tau = t-t_1$. In the weak--coupling limit $d
\overline{\tilde{U}} / dt $ is small and $
\overline{\tilde{U}(t-\tau)}$ may be replaced by
$\overline{\tilde{U}(t)}$. This approximation amounts to neglecting
memory effects and is thus a Markov approximation. We obtain,
\begin{equation}
 \frac{d \overline{\langle nb|\tilde{U}(t)| n'b\rangle}}{dt} = -
 \frac{1}{2}  \sum_{n_1n_2} W_{nn_1n_1n_2}^{(1)} e^{i(E_n-E_{n_2})t} 
  \hspace{0,1cm}\overline{ \langle n_2b|\tilde{U}(t)| n'b\rangle} \ .  
\end{equation}
We have defined
\begin{equation}\label{eqa28}
   W_{nn_1n_1n_2}^{(1)} = 2 \pi \sum_{{b_1}}
  Q_{n{n_1}}Q_{{n_1}{n_2}}  \overline{{V_{b{b_1}}}^2}
  \delta(E_{n_2}\hspace{-0,1cm} + \varepsilon_b\hspace{-0,1cm} -
  E_{n_1} \hspace{-0,1cm}-\varepsilon_{b_1}) \ .
\end{equation}
In the Schr\"odinger picture,
\begin{eqnarray}\label{eqa29}
 \frac{d \overline{\langle nb|U(t)| n'b\rangle}}{dt} =
 &-& i(E_n+\varepsilon_b)  \overline{\langle nb|U(t)| n'b\rangle}
 \nonumber\\
 &-&\frac{1}{2}  \sum_{n_1n_2} W_{nn_1n_1n_2}^{(1)}
  \hspace{0,1cm}\overline{ \langle n_2b|U(t)| n'b\rangle} \ .   
\end{eqnarray}
We have similarly
\begin{eqnarray}\label{eqa30}
 \frac{d \overline{\langle nb|U^\dagger(t)| n'b\rangle}}{dt} =&&
 i(E_{n'}+\varepsilon_b)  \overline{\langle nb|U^\dagger(t)|
   n'b\rangle} \nonumber \\
 &-& \frac{1}{2}  \sum_{n_1n_2} W_{n_2n_1n_1n'}^{(2)}
  \hspace{0,1cm}\overline{ \langle nb|U^\dagger(t)| n_2b\rangle}   
\end{eqnarray}
with 
\begin{equation}\label{eqa31}
 W_{{n_1}n'{n_2}{n_1}}^{(2)} = 2 \pi \sum_{{b_1}}
  Q_{{n_1}n'}Q_{{n_2}{n_1}}  \overline{{V_{b{b_1}}}^2}
  \delta(E_{n_2}\hspace{-0,1cm} + \varepsilon_b\hspace{-0,1cm} -
  E_{n_1} \hspace{-0,1cm}-\varepsilon_{b_1}) \ .
\end{equation}


\subsection{The evolution equation for the average
density operator}

The general density operator at times $t,t'$ is given by
\begin{equation}
  \hat{\rho}(t,t') \theta(t)\theta(t') = K(t) \hat{\rho}(0)K^\dagger(t')   
\end{equation}
where we have used the definition (\ref{eqa1}) for the propagator. We
take the double Fourier transform and obtain
\begin{equation}
 \hat{\rho}(E,E')= G(E) \hat{\rho}(0)G^\dagger(E') 
\end{equation}
where $G(E)$ is given by Eq.~(\ref{eqa4}). Hence
\begin{equation}
\label{eqa11}
  \overline{\rho}(E,E') =
  \sum_{s=0}^{\infty}\sum_{r=0}^{\infty}\overline{G_0(WG_0)^s
  \hat{\rho}(0)(G_0^\dagger W)^rG_0^\dagger} \ .   
\end{equation}
In the weak--coupling limit, $\overline{G}$ appears sandwiched between
two $W$'s which are contracted across $\hat{\rho}(0)$, and the general
term in the expansion (\ref{eqa11}) attains the form
\begin{equation}
  \overline{G} \unitlength0.1cm
  \begin{picture}(41,0)\thinlines 
  \put(2,3.4){\line(0,1){2}}
  \put(11,3.5){\line(0,1){1}}
  \put(2,5.4){\line(1,0){37}}
  \put(0,0){$W$}
  \put(4,0){$\cdots$}
  \put(9,0){$W$}
  \put(13,0){$\overline{G}\hat{\rho}(0)\overline{G}^\dagger$}
  \put(30,3.5){\line(0,1){1}}
  \put(39,3.4){\line(0,1){2}}
  \put(11,4.6){\line(1,0){19}}
  \put(28,0){$W$}
  \put(32,0){$\cdots$}
  \put(37,0){$W$}
  \end{picture}
  \overline{G}^\dagger \ .
\end{equation}
Thus,
\begin{eqnarray}
  \overline{\rho} = \overline{G}\hat{\rho}(0)\overline{G}^\dagger
                         & +& \overline{G} \unitlength0.1cm
  \begin{picture}(23,0)\thinlines 
  \put(2,3.4){\line(0,1){2}}
  \put(2,5.4){\line(1,0){19}}
  \put(0,0){$W$}
  \put(4,0){$\overline{G}\hat{\rho}(0)\overline{G}^\dagger$}
  \put(21,3.4){\line(0,1){2}}
  \put(19,0){$W$}
  \end{picture}
  \overline{G}^\dagger \nonumber  \\
  &+& \overline{G}\unitlength0.1cm
  \begin{picture}(39,0)\thinlines 
  \put(2,3.4){\line(0,1){2}}
  \put(10,3.5){\line(0,1){1}}
  \put(2,5.4){\line(1,0){35}}
  \put(0,0){$W$}
  \put(4,0){$\overline{G}$}
  \put(8,0){$W$}
  \put(12,0){$\overline{G}\hat{\rho}(0)\overline{G}^\dagger$}
  \put(29,3.5){\line(0,1){1}}
  \put(37,3.4){\line(0,1){2}}
  \put(10,4.6){\line(1,0){19}}
  \put(27,0){$W$}
  \put(31,0){$\overline{G}^\dagger$}
  \put(35,0){$W$}
  \end{picture}
\overline{G}^\dagger  + \cdots                         
\end{eqnarray}
or
\begin{equation}
\label{eqa12}
  \overline{\rho} = \overline{G}\hat{\rho}(0)\overline{G}^\dagger +
                              \overline{G}  \unitlength0.1cm
  \begin{picture}(18,0)\thinlines 
  \put(2,3.4){\line(0,1){2}}
  \put(2,5.4){\line(1,0){14}}
  \put(0,0){$W$}
  \put(4,0){$\overline{G}\overline{\rho}\overline{G}^\dagger$}
  \put(16,3.4){\line(0,1){2}}
  \put(14,0){$W$}
  \end{picture}
  \overline{G}^\dagger \ .
\end{equation}
Eq.~(\ref{eqa12}), together with Eq.~(\ref{eqa5}) for $\overline{G}$,
completely determines the dynamics of the system. Transforming back to
the time representation, we can rewrite these equations as 
\begin{eqnarray}
\label{eqa13}
  &&  \overline{\rho}(t,t') \theta(t) \theta(t') = \overline{K}(t)
  \hat{\rho}(0) \overline{K}^\dagger(t') \nonumber \\ 
  & & \hspace{1.5cm}+ \int_{-\infty}^{\infty} d\tau
  \hspace{-0,1cm}\int_{-\infty}^{\infty}   d\tau'
  \overline{K}(t-\tau)
  \unitlength0.1cm
  \begin{picture}(37,0)\thinlines 
  \put(2,3.4){\line(0,1){2}}
  \put(2,5.4){\line(1,0){33}}
  \put(0,0){$W$}
  \put(5,0){$\overline{\rho}(\tau,\tau')
  \theta(\tau) \theta(\tau')$}
  \put(35,3.4){\line(0,1){2}}
  \put(33,0){$W$}
  \end{picture}
  \overline{K}^\dagger(t'-\tau') 
\end{eqnarray}
and
\begin{equation}\label{eqa14}
  \overline{K}(t) = {K}_0 (t)  +  \int_{-\infty}^{\infty}dt_1 
  \int_{-\infty}^{\infty} dt_2  {K}_0 (t-t_1)
   \unitlength0.1cm
  \begin{picture}(30,0)\thinlines 
  \put(2,3.4){\line(0,1){2}}
  \put(2,5.4){\line(1,0){25}}
  \put(0,0){$W$}
  \put(5,0){${K}_0  (t_1-t_2)$}
  \put(27,3.4){\line(0,1){2}}
  \put(25,0){$W$}
  \end{picture}
  \overline{K}(t_2) \ .
\end{equation}
\end{appendix}

\end{document}